\documentclass[aps,prb,reprint,twocolumn,superscriptaddress,showpacs]{revtex4-2}

\usepackage{amsfonts}
\usepackage{mathrsfs}
\usepackage{amsmath}
\usepackage{color}
\usepackage{graphicx}
\usepackage{bm}
\usepackage{amssymb}
\usepackage{xspace}
\usepackage{epstopdf}
\usepackage{dcolumn}
\usepackage{longtable}
\usepackage{multirow}
\usepackage{float}
\usepackage{comment}
\usepackage{tabularx}
\usepackage{dsfont}
\usepackage{fontenc}
\usepackage{verbatim}
\usepackage{color}
\usepackage{natbib}
\usepackage{bm}
\usepackage{amssymb}
\usepackage{xspace}
\usepackage{epstopdf}
\usepackage{dcolumn}
\usepackage{multirow}
\usepackage{tabularx}
\usepackage{bbding}
\usepackage{booktabs}
\usepackage{placeins}
\usepackage[colorlinks=true, letterpaper=true, pdfstartview=FitV, linkcolor=blue, citecolor=blue, urlcolor=blue]{hyperref}

\usepackage[colorlinks=true, letterpaper=true, pdfstartview=FitV, linkcolor=blue, citecolor=blue, urlcolor=blue]{hyperref}

\makeatother

\begin{document}

\title{Weyl-Dirac nodal line phonons with type-selective surface states}

\author{Le Du}
\address{Research Center for Quantum Physics and Technologies, School of Physical Science and Technology, Inner Mongolia University, Hohhot 010021, China}
\address{Key Laboratory of Semiconductor Photovoltaic Technology and Energy Materials at Universities of Inner Mongolia Autonomous Region, Inner Mongolia University, Hohhot 010021, China}

\author{Zeling Li}
\address{Research Center for Quantum Physics and Technologies, School of Physical Science and Technology, Inner Mongolia University, Hohhot 010021, China}
\address{Key Laboratory of Semiconductor Photovoltaic Technology and Energy Materials at Universities of Inner Mongolia Autonomous Region, Inner Mongolia University, Hohhot 010021, China}

\author{Jiabing Chen}
\address{Research Center for Quantum Physics and Technologies, School of Physical Science and Technology, Inner Mongolia University, Hohhot 010021, China}
\address{Key Laboratory of Semiconductor Photovoltaic Technology and Energy Materials at Universities of Inner Mongolia Autonomous Region, Inner Mongolia University, Hohhot 010021, China}

\author{Dongliang Mao}
\address{Research Center for Quantum Physics and Technologies, School of Physical Science and Technology, Inner Mongolia University, Hohhot 010021, China}
\address{Key Laboratory of Semiconductor Photovoltaic Technology and Energy Materials at Universities of Inner Mongolia Autonomous Region, Inner Mongolia University, Hohhot 010021, China}

\author{Lei Wang}
\address{Research Center for Quantum Physics and Technologies, School of Physical Science and Technology, Inner Mongolia University, Hohhot 010021, China}
\address{Inner Mongolia Key Laboratory of Microscale Physics and Atomic Manufacturing, Inner Mongolia University, Hohhot 010021, China}

\author{Xiao-Ping Li}
\email{xpli@imu.edu.cn}
\address{Research Center for Quantum Physics and Technologies, School of Physical Science and Technology, Inner Mongolia University, Hohhot 010021, China}
\address{Key Laboratory of Semiconductor Photovoltaic Technology and Energy Materials at Universities of Inner Mongolia Autonomous Region, Inner Mongolia University, Hohhot 010021, China}

\begin{abstract}
The band complex formed by multiple topological states has attracted extensive attention for the emergent properties produced by the interplay among the constituent states. Here, based on group theory analysis, we present a scheme for rapidly identifying the Weyl-Dirac nodal lines (a complex of Weyl and Dirac nodal lines) in bosonic systems. We find only 5 of the 230 space groups host Weyl-Dirac nodal line phonons. Notably, the Dirac nodal line resides along the high-symmetry line, whereas the Weyl nodal line is distributed on the high-symmetry plane and is interconnected with the Dirac nodal line, jointly forming a composite nodal network structure. Unlike traditional nodal nets, this nodal network exhibits markedly distinct surface states on different surfaces, which can be attributed to the fundamental differences in the topological properties between the Weyl and Dirac nodal lines. This unique property thus allows the material to present distinct surface states in a termination-selective manner. Furthermore, by first-principles calculations,
we identify the materials NdRhO$_3$ and ZnSe$_2$O$_5$ as candidate examples to elaborate the Weyl-Dirac nodal line and their related topological features. Our work provides an insight for exploring and leveraging topological properties in systems with coexisting multiple topological states.
\end{abstract}

\maketitle

\section{Introduction}\label{intro}
The exploration of novel band crossings in crystals and their associated physical properties has attracted significant attention in recent years~\cite{Chiu2016Classification, toposemi2016, bernevig2018recent, weylDirac, yu2022encyclopedia, PhysRevB.105.155156}. According to the dimension of the degeneracy manifold, the band crossings can be classified into three categories, i.e., the zero-dimensional (0D) nodal points represented by Weyl and Dirac points~\cite{murakami2007phase, young2012dirac, wang2012dirac, wang2013three, weng2015weyl, meng2020ternary, PhysRevB.103.L081402, bradlyn2016beyond}, 1D nodal lines (NLs)~\cite{burkov2011topological, kim2015dirac, weng2015topological, fang2016topological, 13li2016dirac, PhysRevB.99.121106, PhysRevB.95.235138, PhysRevB.110.235163, feng2017experimental}, and 2D nodal surfaces~\cite{zhong2016towards, liang2016node, wu2018nodal, zhang2018nodal}. In addition to realizing single topological band crossings, the exploration of multiple topological states coexisting in a single material has recently attracted considerable interest owing to their richer and more diverse topological properties. With multiple symmetries, coexistence can manifest in various forms, such as between nodal points and nodal points~\cite{PhysRevLett.121.106404, PhysRevB.99.075121, PhysRevB.94.165201, PhysRevB.112.155165}, between nodal points and nodal lines~\cite{zhang2017type, PhysRevB.96.045121, PhysRevB.95.235116, PhysRevB.98.035130, PhysRevB.112.045126, PhysRevB.97.241102}, between nodal points and nodal surfaces~\cite{PhysRevB.100.041118}, and between nodal lines and nodal surfaces~\cite{fu2019dirac,PhysRevB.97.235150}.

In contrast to the rich coexistence configurations above, the nodal line-nodal line coexistence is extremely rare due to stricter symmetry required for its formation.
Generally, in the presence of spin-orbital coupling (SOC), the electron system must break inversion ($\mathcal{P}$) or time-reversal symmetry ($\mathcal{T}$) to lift the Kramers degeneracy, enabling the formation of a doubly degenerate Weyl nodal line (WNL). Instead, the Dirac nodal lines (DNLs) always exist with the combination of $\mathcal{PT}$ and extra nonsymmorphic space group symmetries. Therefore, in most cases, these two existence conditions are mutually exclusive and difficult to coexist within the electronic band structure of a single material. Recently, a few studies have pointed out that in systems without SOC, the nonsymmorphic operations composed of glide mirror and screw rotation, together with $\mathcal{T}$, can ensure the coexistence of WNLs and DNLs in crystals with $\mathcal{P}$ symmetry~\cite{PhysRevB.96.155206, PhysRevB.98.075146}. It is suggested that the coexistence of NLs can be realized in light-element materials with negligible SOC. However, such strict conditions significantly narrow the range of candidate materials.

\begin{figure}[t]
	\includegraphics[width=8.4cm]{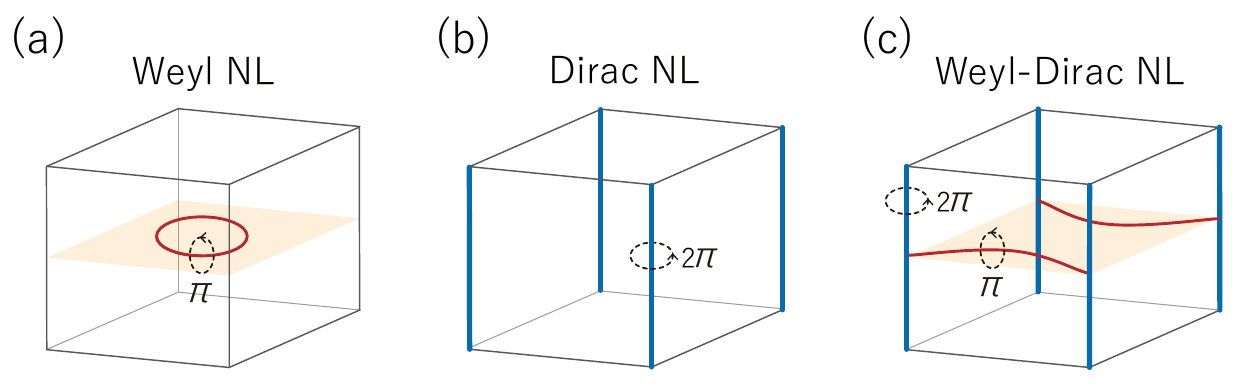}
	\caption{ Schematic figures showing three types of nodal lines (NLs). The orange
region denotes high-symmetry plane. (a) Weyl nodal line (red circle) with a Berry phase of $\pi$. (b) Straight Dirac nodal line (blue solid line) along a high-symmetry line with a Berry phase of $2\pi$. (c) A possible Weyl-Dirac NL structure.
		\label{fig1}}
\end{figure}

Besides electronic systems, phonons, as collective excitations of lattice vibrations in solids, have emerged as an ideal platform for studying novel band degeneracies~\cite{xu2024catalog,li2021computation,fan2024catalog}. Some fascinating phenomena associated with topological phonons in field of  heat transfer, phonon scattering, and electron-phonon interaction are predicted~\cite{RevModPhys.84.1045, liu2020topological, PhysRevMaterials.2.114204}. Compared to electronic bands, phonons, as bosons, are not constrained by the Pauli exclusion principle and Fermi surfaces, allowing all phonon bands to be experimentally probed. Moreover, phonon bands are unaffected by SOC but subject to the same crystal symmetry constraints as electronic bands, offering a broader venue for realizing various quasiparticles and their coexistence~\cite{qin2024diverse, PhysRevB.106.054306, PhysRevB.108.054305, 33b6-1z4b, ft3y-rlcz, PhysRevB.103.104101, PhysRevB.104.085118,  PhysRevLett.124.105303, PhysRevB.109.235415, PhysRevB.101.081403, PhysRevB.108.235302, PhysRevB.110.174111, PhysRevMaterials.5.124203, PhysRevB.104.L041104, PhysRevB.105.174309, PhysRevB.108.115153, PhysRevB.104.174108, PhysRevB.109.155414, fu2023multi, PhysRevB.104.L041107, PhysRevB.108.104312, PhysRevB.108.125127}. Regarding the coexistence of different NL phonons, only specific material has been predicted to host Weyl and Dirac nodal lines~\cite{wang2022hourglass}, and a unified approach for efficiently and rapidly identifying the coexistence of nodal-line phonons, rather than focusing on specific materials, remains lacking. Moreover, since WNLs exhibit a $\pi$ Berry phase [see Fig.~\ref{fig1}(a)] whereas DNLs show $2\pi$ [see Fig.~\ref{fig1}(b)] around a closed loop encircling the line, it remains to be answered whether their coexistence with distinct Berry phases can lead to richer topological phenomena in phononic systems.

In this work, we undertake this task by performing group-theoretical analysis to identify the possible coexistence of Weyl and Dirac NL phonons with distinct degeneracies [see Fig.~\ref{fig1}(c)], referred to as Weyl-Dirac NLs. Our results show that only five space groups  can host such a phonon band complex. Specifically,  the DNLs originate from essential degeneracies along high-symmetry lines (HSLs), while the WNLs reside on high-symmetry planes (HSPL). Through compatibility relation analysis, we identify the formation mechanism of WNLs and their associated band connectivity with DNLs. More importantly, since these two NLs carry different Berry phases and are well-separated in momentum space, the surface signatures of these two types of NLs can be isolated by selecting an appropriate surface projection. This allow for a separate manifestation of WNL-induced drumhead states and DNL-induced torus surface states on distinct surfaces, enabling the realization of type-selective surface states. Based on the first-principles calculations, we identify realistic material candidates, including NdRhO$_{3}$ and ZnSe$_{2}$O$_{5}$. In both materials, the two fascinating surface states arising from Weyl-Dirac NLs can be realized, providing an ideal platform to study the interplay between band crossings with different topological properties.

\section{Approach}

To obtain a complete classification of Weyl-Dirac NLs, which represent a coexistence of the two types of NLs, a natural strategy is to first identify the DNLs and subsequently determine the existence of WNLs among them. For DNLs, a systematic scan was performed on the irreducible representations (IRRs) of the little group at HSLs in the Brillouin zone (BZ) for each of the 230 space groups (SGs). In phonon systems, these IRRs are restricted to single-valued representations, as tabulated in standard references~\cite{bradley2009mathematical}. Specifically, DNLs located along HSLs correspond to four-dimensional IRRs. Based on this criterion, the candidate space groups are identified as Nos. 57, 60, 61, 62, and 205. Moreover, these SGs all possess glide mirror symmetries, which provide the essential symmetry protection for the emergence of WNLs. 

\begin{figure}[t]
\includegraphics[width=8.4cm]{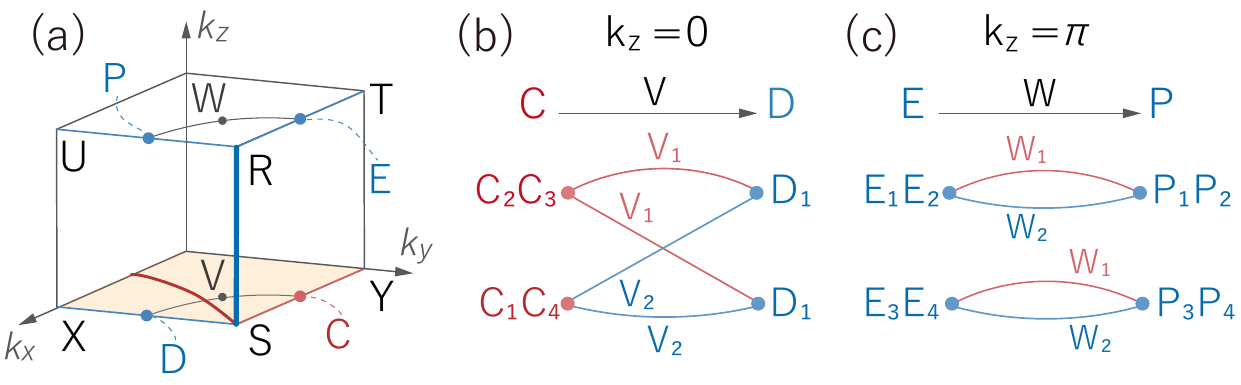}
\caption{Schematic diagram of the WNL formation mechanism. (a) The irreducible BZ for the orthorhombic space group. C, D, E, and P (V and W) denote generic points along high-symmetry lines (on high-symmetry planes). (b)  Schematic energy bands along the C-V-D path in the $k_{z}=0$ plane. (c) Schematic energy bands along the E-W-P path in the $k_{z}=\pi$ plane.  The IRRs $V_{1}$ ($W_{1}$) and $V_{2}$ ($W_{2}$) are indicated by solid lines of different colors (red and blue).
\label{fig2}}
\end{figure}

Without loss of generality, we take SG No. 62 as an example for illustration. The SG 62 belongs to orthorhombic space groups, in which symmetry operators contain three glide mirror symmetries, $\widetilde{M}_{x}=\left\{ M_{x}|\frac{1}{2}\frac{1}{2}\frac{1}{2}\right\} $, $\widetilde{M}_{y}=\left\{ M_{y}|0\frac{1}{2}0\right\} $, and $\widetilde{M}_{z}=\left\{ M_{z}|\frac{1}{2}0\frac{1}{2}\right\} $. Once WNLs appear within HSPLs, these  symmetries ensure the stability of the nodal crossings.
Fig.~\ref{fig2}(a) shows 1/8 of the BZ, and the fourfold degenerate DNL locates at the SR path. Previous studies have indicated that the hourglass-shaped WNL consists of four bands and will cross at a fourfold degenerate point~\cite{PhysRevB.96.155206}. 
Here, fourfold degenerate points are guaranteed to emerge on an HSPL if it is perpendicular to the DNL. Thus, we focus on the $\widetilde{M}_{z}$-invariant planes $k_{z}=0$ and $k_{z}=\pi$, where the DNL intersects to yield the S and R points, respectively.

For the $k_{z}=0$ plane, we focus on the band evolution between two HSLs adjacent to the S point, namely S-Y [labeled $\mathrm{C}\left(u,\frac{1}{2},0\right)$] and S-X [labeled $\mathrm{D}\left(\frac{1}{2},v,0\right)$]. Here and below, $u,v\in(0,\frac{1}{2})$. We define a path that starts from an arbitrary point on C, passes through the in-plane point $\mathrm{V}\left(u,v,0\right)$, and ends at an arbitrary point on D [see Fig.~\ref{fig2}(a)], and we analyze the band connectivity along this path. From a group theoretical point of view, the band connectivity along C-V-D is determined by the compatibility relations between the IRRs of the little group, which can be readily obtained from the decomposition of the representations. In general, the decomposition of a representation can be expressed as~\cite{dresselhaus2007group}
\begin{table}[b]
		\centering
		\caption{Character tables of the little groups at HSPLs  ($k_{z}=0, \pi$) in SG 62 and the decomposition of representations along HSLs into those of the little groups at these planes.}
		\renewcommand{\arraystretch}{1.15}
		\setlength{\tabcolsep}{8.7pt}
		\small
		
		\begin{tabularx}{\columnwidth}{@{} c c c c c c @{}}
			\midrule
			\midrule
			Location  & Irreps & $E$ & $\{ M_z | \tfrac{1}{2} 0 \tfrac{1}{2} \}$ & Decomposition \\
			\midrule
			$k_z=0$ & $V_1$    & 1 & $e^{i\pi u}$    & {} \\
			& $V_2$    & 1 & $-e^{i\pi u}$   & {} \\
			\addlinespace[1pt]
			& $D_1$    & 2 & 0               & $V_1 + V_2$ \\
			& $C_1C_4$ & 2 & $2e^{i\pi u}$   & $2V_1$ \\
			& $C_2C_3$ & 2 & $-2e^{i\pi u}$  & $2V_2$ \\
			\midrule
			$k_z=\pi$ & $W_1$   & 1 & $-e^{i\pi u}$   & {} \\
			& $W_2$   & 1 & $e^{i\pi u}$    & {} \\
			\addlinespace[1pt]
			& $P_1P_2$& 2 & 0               & $W_1 + W_2$ \\
			& $P_3P_4$& 2 & 0               & $W_1 + W_2$ \\
			& $E_1E_2$& 2 & 0               & $W_1 + W_2$ \\
			& $E_3E_4$& 2 & 0               & $W_1 + W_2$ \\
			\midrule
			\midrule
		\end{tabularx}
		\label{table1}
	\end{table}
\begin{equation}
D(R)	=	\sum_{v}c_{v}D^{(v)}(R),
\label{irrs}
\end{equation}
where $R$ denotes the elements of the group, $c_{v}$ represents the multiplicity of the IRR $D^{(v)}$ in the decomposition of $D$,  and is a zero or positive integer. Furthermore, Eq. (\ref{irrs}) can be expressed using the character $\chi$ as
\begin{equation}
\chi(R)	=	\sum_{v}c_{v}\chi^{(v)}(R),
\label{chi}
\end{equation}
and the $c_{v}$ can be obtained via
\begin{equation}
c_{v}	=	\frac{1}{g}\sum_{R}\chi(R)\chi^{(v)}(R)^{*}.
\label{chi}
\end{equation}
Here, $g$ denotes the order of the group. The IRRs of the little group of high-symmetry momentum and the characters of symmetry elements can be obtained from the Bilbao server~\cite{aroyo2011crystallography}. The relevant data for SG 62 are listed in Table~\ref{table1}. It is observed that the HSL C possesses two 2D IRRs, $C_{1}C_{4}$ and $C_{2}C_{3}$, while the D exhibits a 2D IRR, $D_{1}$. The HSPL V contains two 1D IRRs, $V_{1}$ and $V_{2}$. We first examine the decomposition of $D_{1}$ into $V_{1}$ and $V_{2}$. According to Eq.~(\ref{chi}), the decomposition coefficients are determined to be $c_{1}=\frac{1}{2}\times(2\times1+0\times e^{-i\pi u})=1$ and $c_{2}=\frac{1}{2}\times[2\times1+0\times(-e^{-i\pi u})]=1$. Consequently, the compatibility relation between D and V is given by 
\begin{equation}
D_{1}	\rightarrow	V_{1}+V_{2}.
\label{DV}
\end{equation}
Similarly, the compatibility relations between C and V are obtained as
\begin{equation}
C_{1}C_{4}	\rightarrow	2V_{1}, C_{2}C_{3}	\rightarrow	2V_{2}.
\label{CV}
\end{equation}
As a result, typical band structures along the C-V-D path allowed by the compatibility relations are illustrated in Fig.~\ref{fig2}(b). One can find that a band crossing inevitably emerges along the path. This band crossing is symmetry-enforced and protected from gapping owing to the difference in the IRRs $V_{1}$ and $V_{2}$, which have opposite eigenvalues of $\widetilde{M}_{z}$. Since the starting and ending points of the C-V-D path can be chosen arbitrarily, the path itself is arbitrary. For example, selecting an alternative path also yields a band crossing. Thus, such band crossing points are not isolated but form a WNL, as indicated by the red solid curve within the $k_{z}=0$ plane in Fig.~\ref{fig2}(a). We note that such a WNL crosses with the DNL at the S point, and due to the periodicity of the reciprocal space, this gives rise to a composite nodal net.

Regarding the $k_{z}=\pi$ plane,  the path is chosen to start on the $\mathrm{E}(u,\frac{1}{2},\frac{1}{2})$ line, pass through $\mathrm{W}(u,v,\frac{1}{2})$, and end on the $\mathrm{P}(\frac{1}{2},v,\frac{1}{2})$ line [see Fig.~\ref{fig2}(a)]. The decomposition of corresponding IRRs of the little group are listed in Table~\ref{table1}, yielding compatibility relations of 
$E_{1}E_{2}\rightarrow W_{1}+W_{2}$, $E_{3}E_{4}\rightarrow W_{1}+W_{2}$, $P_{1}P_{2}\rightarrow W_{1}+W_{2}$, and $P_{3}P_{4}\rightarrow W_{1}+W_{2}$. Thus, the typical band structures along the E-W-P path are illustrated in Fig.~\ref{fig2}(c). It can be observed that no symmetry-enforced nodal lines exist along this path (while accidental band crossings may occur, which are not discussed here). So far, we have identified the Weyl-Dirac NLs in SG 62 and clarified their formation mechanisms.

\begin{table}[tb]
	\centering
	\caption{Space groups allowing for Weyl-Dirac NLs in phonon systems. Columns 2 and 3 list the locations and irreducible representations (irreps) of DNLs, respectively. Column 4 shows the positions of WNLs in the BZ, column 5 details the compatibility relations (CRs) protecting the WNLs, and column 6 marks the connection points between WNLs and DNLs.}
	\renewcommand{\arraystretch}{1.12}
	\setlength{\tabcolsep}{2pt}
	\scriptsize

	\begin{tabularx}{\columnwidth}{@{}
			c@{\hspace{2pt}}   
			c@{\hspace{2pt}}   
			c@{\hspace{9pt}}  
			c@{\hspace{10pt}}  
			>{\raggedright\arraybackslash}X  
			c                  
			@{}}
		\midrule
		\midrule
		SG & Location & Irreps & Location & \hspace{18pt}CRs & \hspace{-6pt}Connection \\
		& (DNL) &  & (WNL)  &  & \\
		\midrule
		
		\multirow[t]{3}{*}{57 ($Pbcm$)}
		& \multirow[t]{3}{*}{TR}
		& \multirow[t]{3}{*}{$E_1E_1(4)$}
		& \multirow[t]{3}{*}{$k_x=0$}
		& $B_1B_3 \rightarrow 2K_1$
		& T \\
		& & & & $B_2B_4 \rightarrow 2K_2$ & \\
		& & & & $H_1 \rightarrow K_1 + K_2$ & \\
		\addlinespace[1pt]
		
		\multirow[t]{3}{*}{}
		& \multirow[t]{3}{*}{}
		& \multirow[t]{3}{*}{}
		& \multirow[t]{3}{*}{$k_x=\pi$}
		& $P_1P_3 \rightarrow 2L_1$
		& R \\
		& & & & $P_2P_4 \rightarrow 2L_2$ & \\
		& & & & $Q_1 \rightarrow L_1 + L_2$ & \\
		\midrule
		
		\multirow[t]{3}{*}{60 ($Pbcn$)}
		& \multirow[t]{3}{*}{UR}
		& \multirow[t]{3}{*}{$P_1P_1(4)$}
		& \multirow[t]{3}{*}{$k_x=0$}
		& $B_1B_3 \rightarrow 2K_1$
		&  \\
		& & & & $B_2B_4 \rightarrow 2K_2$ & \\
		& & & & $H_1 \rightarrow K_1 + K_2$ & \\
		\addlinespace[1pt]
		
		\multirow[t]{3}{*}{}
		& \multirow[t]{3}{*}{}
		& \multirow[t]{3}{*}{}
		& \multirow[t]{3}{*}{$k_y=0$}
		& $G_1G_4 \rightarrow 2M_1$
		& U \\
		& & & & $G_2G_3 \rightarrow 2M_2$ & \\
		& & & & $A_1 \rightarrow M_1 + M_2$ & \\
		\addlinespace[1pt]
		
		\multirow[t]{3}{*}{}
		& \multirow[t]{3}{*}{}
		& \multirow[t]{3}{*}{}
		& \multirow[t]{3}{*}{$k_y=\pi$}
		& $E_1E_3 \rightarrow 2N_1$
		& R \\
		& & & & $E_2E_4 \rightarrow 2N_2$ & \\
		& & & & $H_1 \rightarrow N_1 + N_2$ & \\
		\midrule
		
		\multirow[t]{3}{*}{61 ($Pbca$)}
		& \multirow[t]{3}{*}{TR}
		& \multirow[t]{3}{*}{$E_1E_1(4)$}
		& \multirow[t]{3}{*}{$k_x=0$}
		& $B_1B_3 \rightarrow 2K_1$
		& T \\
		& & & & $B_2B_4 \rightarrow 2K_2$ & \\
		& & & & $H_1 \rightarrow K_1 + K_2$ & \\
		\addlinespace[1pt]
		
		\multirow[t]{3}{*}{}
		& \multirow[t]{3}{*}{UR}
		& \multirow[t]{3}{*}{$P_1P_1(4)$}
		& \multirow[t]{3}{*}{$k_y=0$}
		& $G_1G_4 \rightarrow 2M_1$
		& U \\
		& & & & $G_2G_3 \rightarrow 2M_2$ & \\
		& & & & $A_1 \rightarrow M_1 + M_2$ & \\
		\addlinespace[1pt]
		
		\multirow[t]{3}{*}{}
		& \multirow[t]{3}{*}{SR}
		& \multirow[t]{3}{*}{$Q_1Q_1(4)$}
		& \multirow[t]{3}{*}{$k_z=0$}
		& $C_1C_4 \rightarrow 2V_1$
		& S \\
		& & & & $C_2C_3 \rightarrow 2V_2$ & \\
		& & & & $D_1 \rightarrow V_1 + V_2$ & \\
		\midrule
		
		\multirow[t]{3}{*}{62 ($Pnma$)}
		& \multirow[t]{3}{*}{SR}
		& \multirow[t]{3}{*}{$Q_1Q_1(4)$}
		& \multirow[t]{3}{*}{$k_z=0$}
		& $C_1C_4 \rightarrow 2V_1$
		& S \\
		& & & & $C_2C_3 \rightarrow 2V_2$ & \\
		& & & & $D_1 \rightarrow V_1 + V_2$ & \\
		\midrule
		
		\multirow[t]{3}{*}{205 ($Pa$-3)}
		& \multirow[t]{3}{*}{MR}
		& \multirow[t]{3}{*}{$T_1T_1(4)$}
		& \multirow[t]{3}{*}{$k_z=0$}
		& $Z_1Z_4 \rightarrow 2A_1$
		& M \\
		& & & & $Z_2Z_3 \rightarrow 2A_2$ & \\
		& & & & $ZA_1 \rightarrow A_1 + A_2$ & \\
		\midrule
		\midrule
	\end{tabularx}
	\label{table2}
\end{table}

The analysis for other SGs proceeds in a similar way, and the result is presented in Table~\ref{table2}. For 5 candidate SGs hosting Weyl-Dirac NLs, the locations of the DNLs and their corresponding IRRs are presented, together with the planes hosting the symmetry-enforced WNLs and the compatibility relations stabilizing them (Note that the naming convention for wave vectors of the Bilbao Server is strictly followed~\cite{aroyo2011crystallography}). The connection points between the Dirac and Weyl NLs are also listed. Note that in addition to the DNL existing along the UR path in SG 60, there also exists a fourfold degenerate Dirac point at the T point. As a result, an hourglass-shaped WNL can exist under the allowed compatibility relation, but is not connected to the DNL. By contrast, the WNLs lying in the $k_{y}=0$ and $\pi$ planes maintain strict connectivity with the DNL.

\section{Material candidates}
In this section, we present two material candidates that host Weyl-Dirac NL phonons and investigate their rich surface states. These materials serve as an ideal platform for exploring the interplay between distinct topological states.
\begin{figure}[t]
	\includegraphics[width=8.4cm]{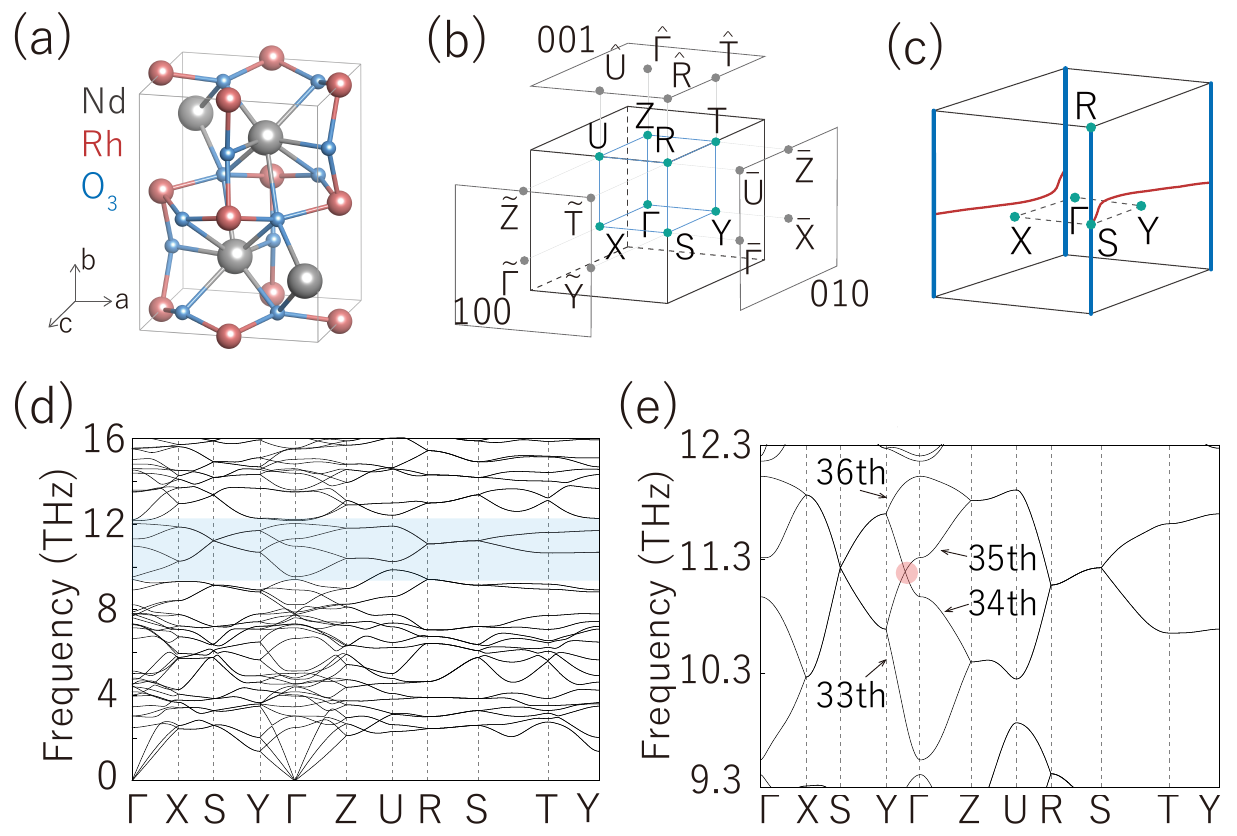}
	\caption{(a) Crystal structure of NdRhO$_3$. (b) Bulk and surface BZs. (c) Schematic showing Dirac-Weyl NL. The Dirac (Weyl) NL is indicated by blue (red) solid line. (d) Phonon spectrum of NdRhO$_3$ along high-symmetry paths. (e) Enlarged phonon dispersions of blue regions in (d). The red dots mark the band crossings forming the WNLs.
		\label{fig3}}
\end{figure}
\subsection{\label{Msymm} Example 1: NdRhO$_{3}$ }
The first example is NdRhO$_{3}$, which has been experimentally synthesized~\cite{macquart2006crystal, yi2013crystal}. Its crystal structure and BZ are illustrated in Figs.~\ref{fig3}(a) and \ref{fig3}(b). NdRhO$_{3}$ belongs to SG 62 ($Pnma$), one of the candidates in Table~\ref{table2}. 

We performed first-principles calculations based on density functional theory (DFT) using the Vienna \textit{Ab-initio} Simulation Package (VASP)~\cite{PhysRevB.49.14251,PhysRevB.54.11169, PhysRevB.50.17953}. The exchange-correlation potential was treated within the generalized gradient approximation (GGA) with Perdew-Burke-Ernzerhof (PBE)~\cite{PhysRevLett.77.3865}. The plane-wave cutoff energy was set to 520 eV, and a $\Gamma$-centered $k$-mesh of $11 \times 12 \times 11$ was used for Brillouin zone sampling. Convergence criteria were set to $10^{-8}$ eV for energy and 0.01 eV/\text{\AA} for forces. Phonon spectra were calculated based on density functional perturbation theory (DFPT)~\cite{RevModPhys.73.515} with a $2 \times 1 \times 2$ supercell, and the force constants matrices were constructed via the PHONOPY code~\cite{togo2015first}. The topological properties were calculated using the iterative Green’s function method based on the phononic tight-binding Hamiltonian constructed with the WannierTools software package~\cite{wu2018wanniertools}

According to Table~\ref{table2}, SG 62 will host a DNL at SR lines and WNLs in $k_{z}=0$ plane. In Fig.~\ref{fig3}(d), we plot the phonon band structure for NdRhO$_{3}$,  with a zoomed-in view of the 9.3-12.3 THz range shown in Fig.~\ref{fig3}(e). One observes that there is indeed a fourfold-degenerate DNL along the SR path, consisting of phonon branches 33 to 36. Furthermore, one can find a two-band crossing between bands 34 and 35 along the $\Gamma$-Y path. This point is not isolated due to the presence of $\widetilde{M}_{z}$ symmetry. A careful scan shows that it resides on a WNL in the $k_{z}=0$ plane, as schematically shown in Fig.~\ref{fig3}(c). Moreover, the intersection of the Weyl and Dirac nodal lines at point S results in a Weyl-Dirac nodal net, as predicted by the group theory analysis in Table~\ref{table2}.
\begin{figure}[t]
	\includegraphics[width=8.4cm]{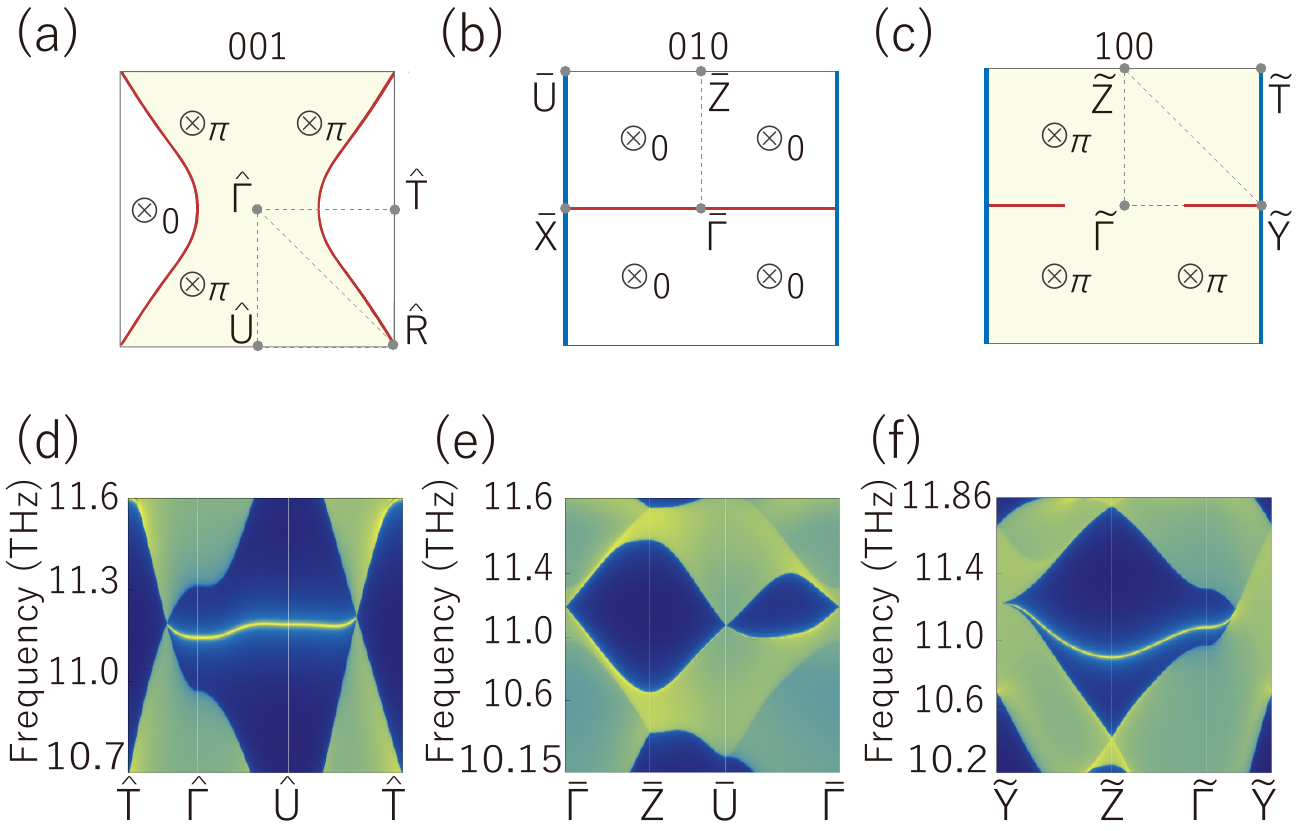}
	\caption{The surface properties of NdRhO$_3$. (a)-(c) Schematics of the surface states for the (001), (010), and (100) surfaces. Red and blue lines indicate projections of bulk WNLs and DNLs onto the surface BZ, respectively. The values 0 and $\pi$ represent the Zak phases for lines normal to each surface. (d)-(f) Calculated surface spectra on the (001), (010), and (100) surfaces. (d) shows the drumhead surface states, while (f) shows the TSS.
		\label{fig4}}
\end{figure}

Next, we investigate the surface modes associated with Weyl-Dirac NL state in NdRhO$_{3}$. Owing to the difference in degeneracy,  the WNLs and DNLs exhibit Berry phases of $\pi$ and $2\pi$, respectively, for a closed loop enclosing the NLs. Such a distinction may result in the Weyl-Dirac NLs exhibiting multiple types of surface states, fundamentally different from conventional single nodal line systems. First, we consider the (001) surface. The surface projections of the WNLs are denoted by red solid lines, while the DNL (SR) projects to a single point $\hat{R}$. Therefore, the surface properties of the (001) surface are dominated by the WNL, with negligible DNL contribution. We then calculate the Zak phase in the bulk BZ along a straight line transverse to the (001) surface and plot the corresponding surface states [see Figs.~\ref{fig4}(a) and (d)], which are typical drumhead-like surface states induced by the WNL. Next, we turn our attention to the (100) and (010) surfaces, where the surface states are primarily dominated by the DNL. Unlike the WNL shown in Fig.~\ref{fig4}(a), the DNL does not separate the surface BZ into two parts, and the whole surface BZ would share the same topological properties [see Figs.~\ref{fig4}(b) and (c)]. Similar calculations of the Zak phase for the (100) and (010) surfaces are presented in Figs.~\ref{fig4}(c) and (b). The results show it is $\pi$ for the (100) surface and 0 for the (010) surface. Consequently, the (100) surface hosts torus surface states (TSS)~\cite{PhysRevB.99.121106, PhysRevB.103.L161109} that span the entire surface BZ [see Fig.~\ref{fig4}(f)], while the (010) surface exhibits no surface states [see Fig.~\ref{fig4}(e)] (Note that the projections of the two WNLs overlap on the (010) surface, leaving the Zak phase unchanged despite the surface BZ being partitioned into two parts). Therefore, the Weyl-Dirac NLs in NdRhO$_{3}$ can exhibit distinct types of surface modes depending on the surface orientation, thereby realizing type-selective surface states in a phononic system.
\begin{figure}[t]
	\includegraphics[width=8.4cm]{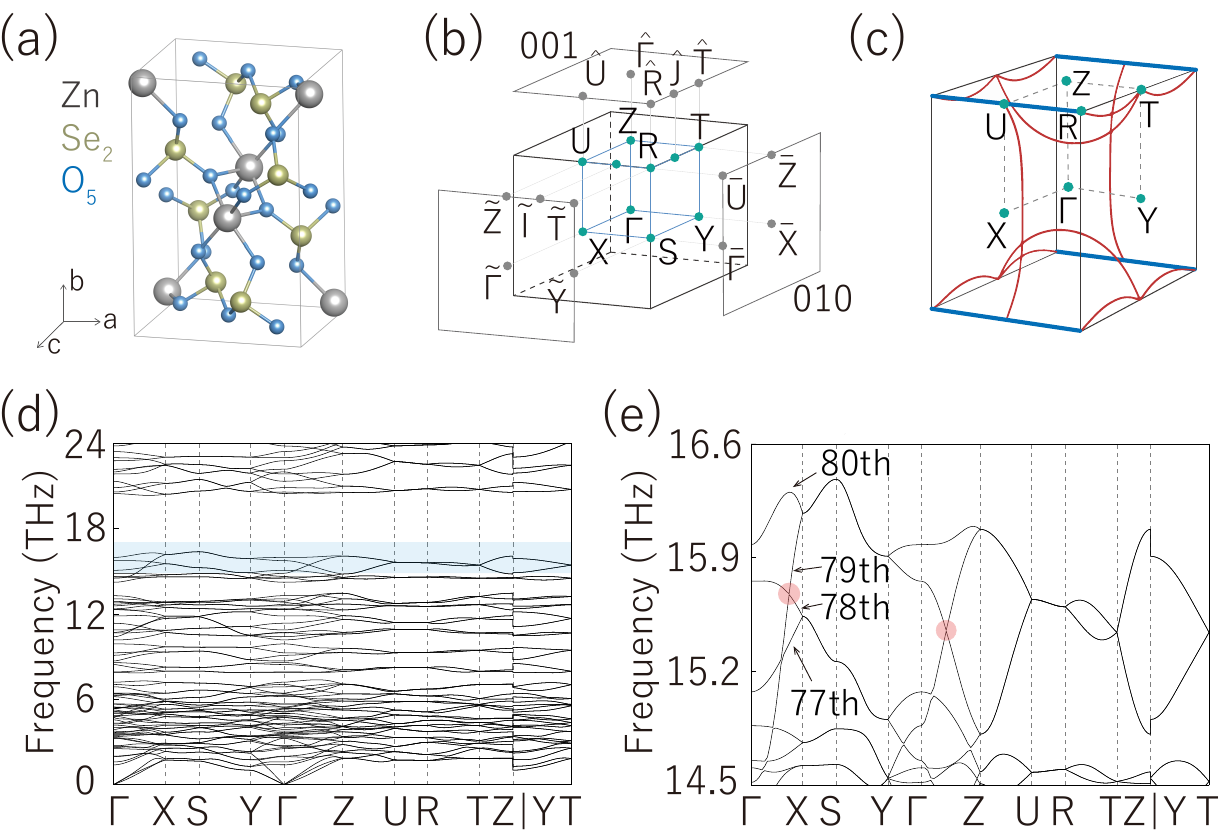}
	\caption{(a) Crystal structure of ZnSe$_2$O$_5$ material. (b) Bulk and surface BZs. (c) Schematic illustration of the Dirac (Weyl) nodal lines, indicated by blue (red) solid lines. (d) Phonon dispersion of ZnSe$_2$O$_5$ along high-symmetry paths. (e) Enlarged view of the phonon dispersion in the blue regions of (d). Red dots mark the band crossings forming the WNLs.
		\label{fig5}}
\end{figure}

\subsection{\label{Msymm} Example 2: ZnSe$_{2}$O$_{3}$ }
The second example is the phonons in ZnSe$_{2}$O$_{3}$, which has also been successfully synthesized experimentally~\cite{meunier1974cristallochimie}. The material belongs to SG 60 ($Pbcn$), and its lattice structure and BZ are illustrated in Figs.~\ref{fig5}(a) and (b).

According to Table~\ref{table2}, SG 60 can host the DNL along UR path and WNLs on the $k_{x}=0$, $k_{y}=0$, and $k_{y}=\pi$ planes. In Fig.~\ref{fig5}(d), we plot the phonon band structure of ZnSe$_{2}$O$_{3}$ and provide a zoomed-in view of the 14.5-16.6 THz region in Fig.~\ref{fig5}(e). One can find that there indeed exist a DNL on UR path. Moreover, an isolated fourfold-degenerate Dirac point emerges at the T point. As discussed previously, such an extra fourfold-degenerate point in addition to DNL is a unique feature of SG 60. In addition, another two band crossings are found along the $\Gamma$-X and $\Gamma$-Z paths, formed by the crossing of phonon branches 78 and 79. These crossing points are not isolated due to the presence of $\widetilde{M}_{y}$ and $\widetilde{M}_{x}$ in SG 60. 
We then perform a careful scan and result is shown as red solid lines in Fig.~\ref{fig5}(c). One observes that there indeed exist WNLs in the $k_{x}=0$ and $k_{y}=0, \pi$ planes, consistent with symmetry analysis (see Table~\ref{table2}). It should be noted that these WNLs exhibit connectivity not only with the DNLs but also with the four-fold degenerate point T, together forming the Weyl-Dirac nodal network.
\begin{figure}[t]
	\includegraphics[width=8.4cm]{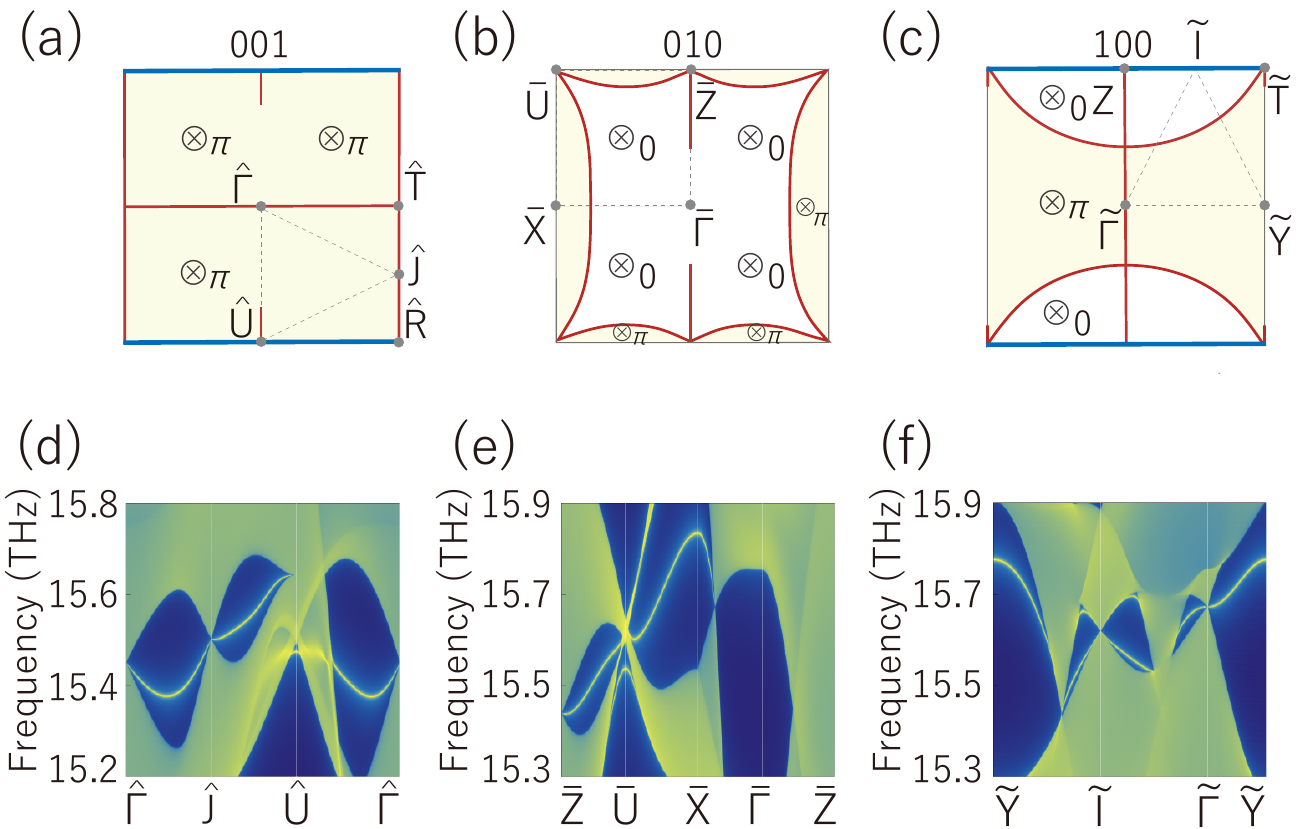}
	\caption{The surface properties of ZnSe$_{2}$O$_5$. (a)-(c) Schematic illustrations of the surface states for the (001), (010), and (100) surfaces. The red and blue lines indicate the projections of the bulk WNLs and DNLs, respectively. The values 0 and $\pi$ denote the Zak phases calculated for lines perpendicular to each surface. (d)-(f) Calculated surface dispersions for the (001), (010), and (100) surfaces.
		\label{fig6}}
\end{figure}

In Figs.~\ref{fig6}(a)-(c), we plot the projections of the Weyl and Dirac NLs onto the (001), (010), and (100) surface BZ, respectively, alongside the calculated Zak phases for distinct regions. The pale yellow areas in the surface BZ denote a Zak phase of $\pi$, indicating the presence of surface states. As a result, similar argument shows that the (001) surface of ZnSe$_{2}$O$_{3}$ exhibits TSS [see Fig.~\ref{fig6}(d)], whereas the (010) surface hosts drumhead-like surface states [see Fig.~\ref{fig6}(e)]. Regarding the (100) surface, while Dirac and Weyl NLs coexist in the projection, the surface states are primarily dominated by the WNL topology. This is because the projection of the WNL in the $k_{x}=0$ plane partitions the surface BZ into two distinct regions. When traversing the (100) surface along a straight line, the Zak phase changes upon crossing the WNL but remains invariant upon crossing the DNL, resulting in drumhead-like surface states [Fig.~\ref{fig6}(f)]. 

\section{\label{discussion} Conclusion and Discussion}
In conclusion, we propose a rational approach for the exhaustive exploration of the Weyl-Dirac NL phonons along HSLs and related HSPLs of the BZ across all space groups. As discussed, employing group-theoretical analysis, we not only determine the DNL configurations but also derive the compatibility relations enforcing the emergence of WNLs within the 5 candidate space groups. Since the DNL corresponds to a unique IRRs, these Weyl-Dirac NLs can exist across the entire frequency spectrum. This characteristic is highly advantageous for their exploration and identification in material systems. Furthermore, we have predicted two material candidates,  NdRhO$_{3}$ and ZnSe$_{2}$O$_{3}$, with ideal Weyl-Dirac NL states in their phonon bands. Interestingly, we revealed that Weyl-Dirac NLs exhibit type-selective topological surface states, displaying either drumhead-like states or unique TSS depending on the chosen surface projection. Notably, since multiple WNLs lead to complex surface state configurations, we prioritize SGs 62 and 205 as ideal candidates for material realization. As indicated in Table~\ref{table2}, these two groups are distinguished by hosting WNLs in only one high-symmetry plane, thereby yielding clean surface states essential for effective surface state engineering. 

\begin{acknowledgments}
 This work is supported by the National Natural Science Foundation of China (Grants No. 12564017, No. 12304086, and No. 12304165).
\end{acknowledgments}



\bibliography{ref}

\end{document}